\documentclass [12pt] {scrartcl}
\usepackage{blindtext}
\usepackage{multirow}
\usepackage{graphicx}

\usepackage[natbibapa]{apacite}
\usepackage{xfrac}
\usepackage{bbm}
\usepackage{slashed}
\usepackage{mathtools}
\usepackage{dsfont}

\usepackage{amsthm, amsmath, array}
\numberwithin{equation}{section}

\title{\Large Motivating Gauge-Invariant Approaches to Particle Physics}
\author{P. Berghofer${\,}^{*}$}
\date{}

\begin{document}

\maketitle

\begin{center}
\vskip -0.8cm
\noindent

${}^*$ Department of Philosophy -- University of Graz \\
Heinrichstraße 26/5, 8010 Graz, Austria \\
philipp.berghofer@uni-graz.at

\end{center}

\section*{Abstract}

There is noticeable consensus among physicists and philosophers that only gauge-invariant quantities can be physically real. However, this insight that physical quantities must be gauge-invariant is not well-reflected in standard approaches to particle physics. For instance, each and every elementary field/particle of the Standard Model fails to be gauge-invariant! The main objective of this paper is to offer an accessible, concise, and convincing analysis of why philosophers and physicists should devote more of their energy to working on gauge-invariant approaches. Correspondingly, the thesis of this paper is that pursuing gauge-invariant approaches has several virtues. For instance, gauge-invariant reformulations allow us to make particle physics consistent with the mathematical framework in which it is formulated. This is illustrated by how mathematical theorems such as Elitzur’s theorem, the Gribov ambiguity, and Haag’s theorem pose problems for standard approaches but are avoided by gauge-invariant approaches.

\section{Introduction}

The Standard Model of particle physics is arguably the greatest triumph of science to date. In terms of agreement between theoretical prediction and empirical confirmation it is unrivaled. When we think of the fundamental building blocks of reality (according to our currently best physics), it is natural to identify them as the elementary fields/particles of the Standard Model, such as the electron field or the electromagnetic field. The Standard Model is a quantum field theory. More precisely, it is a quantum \textit{gauge} field theory. That is, the theory rests on internal \textit{local} symmetries. There is much debate in the philosophy literature on how to interpret gauge symmetries. Importantly, there is a noticeable consensus among philosophers and physicists that only gauge-invariant quantities can be physically real. But this insight that physical quantities must be gauge-invariant is not at all reflected in the standard approaches to physics. In particular, each and every “elementary” field/particle\footnote{Note that here by “particles” I don’t mean the experimentally detected physical objects (which I consider to be properly mathematically represented only by gauge-invariant quantities) but the “formal” quanta of the gauge-variant bare Lagrangian fields.} in the Standard Model is actually a non-gauge-invariant quantity.\footnote{With the possible exception of right-handed neutrinos which are formally gauge-invariant but experimentally unconfirmed. In what follows, I don’t consider right-handed neutrinos as they typically are not viewed as a proper part of the Standard Model.}

Furthermore, physical mechanisms are often explained in terms of gauge-variant quantities. For instance, textbook approaches to the BEH mechanism imply that the Higgs field gives mass to particles via the spontaneous breaking of a gauge symmetry. On a conceptual level, this seems problematic as it is unclear how the breaking of unphysical mathematical redundancy (as gauge symmetries are often interpreted) could have such an impact on the physical world. More importantly for the purposes of this paper, on a technical level, there is Elitzur’s theorem that says that gauge symmetries cannot be broken spontaneously. While Elitzur’s theorem is not often discussed in philosophy of physics, Haag’s theorem has become very prominent. Among physicists, Haag’s theorem is known to be very “inconvenient” as it implies that perturbation theory is ill-defined. However, since perturbation theory is tremendously successful, it is an open problem to understand how perturbation theory can be both mathematically forbidden and empirically successful. In this paper, it will be argued that pursuing gauge-invariant approaches allows us to avoid all these problems. The BEH mechanism can be reformulated in a gauge-invariant fashion such that Elitzur’s theorem is not violated and furthermore this allows us not only to avoid Haag’s theorem but to explain the success of perturbation theory. 

The main objective of this paper is to offer an accessible, concise, and convincing analysis of why philosophers and physicists should devote more of their energy to working on gauge-invariant approaches.\footnote{Of course, the main (and perhaps only truly compelling) motivation in favor of gauge-invariant approaches would come in the form of empirical evidence. That is, it must be shown that gauge-invariant approaches and standard perturbative approaches differ in their empirical implications. Importantly, as addressed below, all gauge-invariant approaches imply that the fundamental observable building blocks of reality are not point-like particles but rather extended composite objects, and it has been argued that it will be possible to empirically test this implication in the not-too-distant future. In particular, the so-called Fröhlich-Morchio-Strocchi mechanism is expected to have empirical implications that can be tested against standard approaches to the BEH effect. While this paper cannot contribute to this research, its objective is to motivate why research in this direction should be intensified. More specifically, I argue that gauge-invariant approaches have virtues that so far have not been addressed in the philosophy of physics literature, such as avoiding the problems raised by Haag’s theorem and explaining the success of perturbation theory.} Correspondingly, the thesis of this paper is that pursuing gauge-invariant approaches has several virtues. Section 2 summarizes why there is this consensus among physicists and philosophers that physical quantities must be gauge-invariant and shows that this insight is not consistently reflected in our standard Lagrangian theories. We will see how quantities such as the electron field and the electromagnetic field can be “made” gauge-invariant and discuss ontological implications. In Section 3, we address one of the main virtues of gauge-invariant approaches that have so far been ignored in the philosophy literature, namely how this allows us to make particle physics consistent with the mathematical framework in which it is formulated. Our focus here will be on Haag’s theorem. In Section 4, we will see how a gauge-invariant reformulation of the electron, as illustrated in Section 2, allows us to explain a physical phenomenon, namely the Aharonov-Bohm effect.

\section{Initial motivation and ontological implications}

In this section, I briefly outline why there is consensus that only gauge-invariant quantities can be physically real. Then I show that this insight that physical quantities must be gauge-invariant is completely ignored in standard approaches to particle physics. Each and every elementary field/particle of the Standard Model fails to be gauge-invariant. I use QED as an example to discuss how we can reformulate the respective quantities in a gauge-invariant fashion and point toward some ontological implications.

In classical electrodynamics, the Lagrangian (in the absence of any sources) takes the form

\begin{equation}
\label{LCED}
{\cal L}=-\frac{1}{4}F_{\mu \nu}^2 .
\end{equation}

This is a physically central quantity as it is the Lagrangian of massless spin-1 fields. $F_{\mu \nu}$ is the field-strength tensor, which can be expressed as $$ F_{\mu \nu} = \partial_\mu A_\nu - \partial_\nu A_\mu  .   $$

Here $\partial_\mu$ is the four-gradient and  $A_\mu$ is the gauge field, namely the electromagnetic four-potential. Accordingly, the Lagrangian can be written as 

$${\cal L}=-\frac{1}{4}(\partial_\mu A_\nu - \partial_\nu A_\mu)^2. $$

As is well-known, classical electrodynamics is a gauge theory; in fact, it is the paradigm gauge-theory. The gauge transformation of the theory is

\begin{equation}
\label{transfo}
 A_\mu (x) \rightarrow A_\mu (x) + \partial_\mu f(x),
\end{equation}

 $ f(x)$ being an arbitrary function. Since $f(x)$ depends on position, (\ref{transfo}) is a \textit{local} transformation. Classical electrodynamics is a gauge theory precisely because the Lagrangian is invariant under (\ref{transfo}). The electromagnetic four-potential, of course, is \textit{not} gauge-invariant. Infinitely many different $A_\mu$ lead to the same Lagrangian.

Now, what are the physical quantities of classical electrodynamics? Obviously, the gauge field does not suggest itself, as it can be chosen arbitrarily. Importantly, from the gauge field the electric and the magnetic field can be derived as

\begin{equation}
\label{EM}
\begin{aligned}
E_i = \partial_0 A_i - \partial_i A_0, \\
B_i = \epsilon_{ijk} \partial_j A_k .
\end{aligned}
\end{equation}

While Greek indices run over the space-time components (0-3), Latin indices run only over the spatial components (1-3). The electric and the magnetic field are gauge-invariant quantities, i.e., they remain invariant under the gauge transformation  (\ref{transfo}). Unsurprisingly, there is basically universal agreement that in classical electrodynamics the physical quantities are the electric and the magnetic field. These are the quantities that remain invariant and can be observed. The gauge field, by contrast, is widely considered to be unphysical. It is a mathematical tool, highly convenient in many calculations but devoid of physical content (see, e.g., \citealp[95]{Anderson1967}; \citealp[221]{Duncan2012}; \citealp[3]{HT1992}; \citealp[§18]{LL1994}; \citealp[130]{Schwartz2014}; \citealp[169]{Tong2016}; \citealp[187]{Zee2010}; \citealp[546]{ZinnJustin}).

Given the above, it does not come as a surprise that there is considerable consensus among physicists and philosophers that only gauge-invariant quantities can be physically real (for an overview, see \citealp{Berghofer-et-al2023}). However, what is surprising, in this light, is that each and every ``elementary" field in the Standard Model of particle physics is actually a gauge-variant, i.e., a non-gauge-invariant, quantity! In what follows, we illustrate and discuss this in the context of QED. The Lagrangian for QED reads

\begin{equation}
\label{QED}
{\cal L_{QED}}=-\frac{1}{4}F_{\mu \nu}^2 + i \bar\psi \gamma^\mu \partial_\mu \psi - e \bar\psi \gamma^\mu A_\mu \psi  - m \bar\psi  \psi         .     
\end{equation}

Here $\psi$ is a spinor field, namely the electron-positron field. $A_\mu$ is the gauge field, referred to as the photon field or (not entirely accurately) the electromagnetic field. $\gamma^\mu $ are Dirac matrices and $m$ is the mass of the electron or positron. Now, importantly, ${\cal L_{QED}}$ is invariant under the following gauge transformations

\begin{align}
 A_\mu (x) \rightarrow A_\mu (x) + \partial_\mu f(x) && \psi (x) \rightarrow e^{-i e f(x)} \psi (x). 
\end{align}

So, what are the physical quantities of QED? The standard answer is to say that the electon field, $\psi$, and the photon field, $A_\mu$, are the basic physical objects. In this picture, the electron field and the photon field are fundamental and each and every electron/photon can be understood as an excitation of the respective underlying quantum field.\footnote{As Steven Weinberg put it: ``Just as there is an electromagnetic field, whose energy and momentum come in tiny bundles called photons, so there is also an electron field, whose energy and momentum and electric charge are found in the bundles we call electrons, and likewise for every species of elementary particle. The basic ingredients of nature are fields; particles are derivative phenomena" (\citealp[221]{Weinberg2001}). The other main interpretation of QFT is to say that not fields but particles are the fundamental objects, mathematically represented by the creation operators of the respective quantum field. For an overview on this debate, see (\citealp{Baker2016}). Importantly, both interpretations imply that the fundamental physical objects are \textit{not} gauge-invariant quantitities!} But the electron field and the photon field are \textit{not} gauge-invariant! Just as in the case of classical electrodynamics, infinitely many mathematically distinct gauge-related spinor/vector fields lead to the same Lagrangian.

Sure, physicists have developed powerful methods, such as gauge fixing, to reduce gauge symmetries and make successful calculations. But the question remains: What are the physical quantities of QED (or the Standard Model in general)? If we rule out the ``electron" field, $\psi$, and the ``photon" field, $A_\mu$, as they are not gauge-invariant, where does this leave us? In classical electrodynamics, we pursued as follows: We performed mathematical operations on the gauge-variant fields to build new gauge-invariant fields (\ref{EM}):
\begin{align*}
E_i = \partial_0 A_i - \partial_i A_0, \\
B_i = \epsilon_{ijk} \partial_j A_k .
\end{align*}

Now, of course, the question is whether we can do something similar in QED. The answer is yes. More precisely, we can ``convert" the gauge field into a gauge-invariant field by transverse-projecting the gauge field (\citealp[139]{Maas2019b}; see also \citealp[Section 80]{Dirac1958} and \citealp{McMullan-Lavelle1995,McMullan-Lavelle1997}):

\begin{equation}
 A_\mu^P = \Big(g_{\mu \nu} - \frac{p_\mu p_\nu }{p^2}\Big) A^\nu .
\end{equation}

This physical photon field, $ A_\mu^P $, is invariant under the gauge transformation $ A_\mu  \rightarrow A_\mu + \partial_\mu f .$ Fourier transforming back to position space (\citealp[equ. 7.9]{Maas2019b}), 
\begin{equation}
\label{physicalp}
A_\mu^P (x) = g_{\mu \nu} A^\nu (x) - \frac{1}{4 \pi} \int d^4 y \frac{\partial_\mu \partial_\nu}{|x-y|} A^\nu (y). 
\end{equation}

In what follows, we are particularly interested in the matter field of QED, i.e., the electron field. The gauge-variant $\psi$ can be ``converted" into a gauge-invariant quantity by performing the following operation (\citealp[140, equ. 7.12]{Maas2019b}):

\begin{equation}
\label{physicale}
\psi^P (x) = D(x)\psi(x),
\end{equation}
$D(x)$ being the Dirac phase factor
\begin{equation}
D(x) = \mathrm{exp} \bigg(-ie \int_x ^\infty dy_\mu A_\mu (y)\bigg).
\end{equation}

This physical electron field is also a gauge-invariant quantity.
\\~\\
The first to explictly argue that the physical electron takes the form (\ref{physicale}) was Paul Dirac (1955, equ. 42). Since then, a number of physicists have pursued such a manifestly gauge-invariant approach (see, in particular, \citealp{McMullan-Lavelle1995,McMullan-Lavelle1997} and \citealp{Lavelle-McMullan-Bagan2000a,Lavelle-McMullan-Bagan2000b}). It is beyond doubt that if such a manifestly gauge-invariant approach is on the right track, this has massive ontological-philosophical implications regarding the nature of the fundamental building blocks of reality. One immediate implication, emphasized by Dirac, is that we cannot conceptualize the electron as separated from its surrounding cloud of photons. ``\textit{A theory that works entirely with gauge-invariant operators has its electrons and positrons always accompanied by Coulomb fields around them}, which is very reasonable from the physical point of view" (\citealp[657]{Dirac1955}). Similarly, Lavelle and McMullan have summarized this as follows:
\begin{quote}
In a recent series of papers [...] we have investigated the physical degrees of freedom in abelian and non-abelian gauge theories and the intimately related question of gauge fixing. For Quantum Electrodynamics (QED) the physical degrees of freedom are the two transverse photon polarisations and the observed electron. This electron is not the Lagrangian fermion, which is neither gauge invariant nor associated with an electric field. In fact the physical electron is this fermion accompanied by a non-local photonic cloud. (\citealp[89]{McMullan-Lavelle1995})
\end{quote}

In other words:

\begin{quote}
This directly implies that the fundamental matter fields in the original Lagrangian of QED should not be identified with the asymptotic physical fields. In particular all charged fields, whether quarks, gluons, or electrons, must always carry with them a chromo-or electromagnetic cloud and only these systems - the matter and its associated dressing cloud taken together - can have any physical meaning. (\citealp[527]{Lavelle-McMullan-Bagan2000b})
\end{quote}

Before pondering possible ontological implications, let us first get clear on the terminology. In this picture, the ``Lagrangian" or ``bare" electron field is the gauge-variant $\psi$ and the physical electron field is  $ \psi^P$ as it is defined in (\ref{physicale}).\footnote{It is to be noted that I do not want to suggest that there is a unique way of dressing the electron. That is, I do not claim that the gauge-invariant physical electron can only be understood according to the dressing as defined in (\ref{physicale}). In fact, there are infinitely many mathematically different ways to dress the electron (see, e.g., \cite[7]{Giddings-Weinberg2020}. Importantly, any such dressing (at least any known to the literature) “is a functional of $A_\mu$” \cite[7]{Giddings-Weinberg2020}, which means that the electron is dressed up with photons. It it also to be noted that there is a close and subtle relationship between dressings and gauge fixings and that to every gauge fixing there corresponds a dressing (see \cite{McMullan-Lavelle1997} and \cite{Gomes2024}). As argued by Gomes, this can be interpreted as undermining the idea that dressings lead to a clear ontological picture. Such discussions are further complicated by the fact that the distinctions between dressings and fixings are mathematically very subtle to the effect that what are often thought as prominent examples of fixings actually are dressings (see \cite{BerghoferFrancois} and \cite{Guillaud}). Personally, my hope/expectation is that even if there are mathematically distinct dressings available, physically we get a clear picture, such as electrons dressed with photons in QED.} Correspondingly, the ``Lagrangian" or ``bare" electron is an excitation of  $\psi$ and the physical electron is an excitation of $ \psi^P$. According to the interpretation of Dirac, Bagan, Lavelle, and McMullan, only the gauge-invariant $ \psi^P$ is physically real, while the gauge-variant $\psi$ is not. The physical electron is understood as a ``composite" object consisting of an elementary electron being surrounded by a cloud of elementary photons. Prima facie, this seems to be paradoxical on two fronts. First, how can a fundamental object be a composite object? Second, how can a physical object consist of non-physical ones? The paradoxicality of this situation is clearly visible when McMullan and Lavelle in the above quotation say that the physical electron is not the Lagrangian fermion but the Lagrangian fermion accompanied by a photonic cloud or when Bagan et al. say that the ``matter" fields should not be identified with the physical fields.

I take it that it is obvious that this deserves detailed philosophical reflection. One way of trying to make sense of this is to say that the underlying quantum field(s) cannot be excited in a way such that a single bare electron exists in the universe. Instead, every electron in the universe is surrounded by a cloud of photons. In this sense, the bare Lagrangian electron must be distinguished from a physical electron. But every physical electron is a bare electron surrounded by photons. In this sense, bare electrons do exist. But they never exist as free point-like objects but only as being a constituent of a composite extended object.

Quarks are a helpful analogy. Single Lagrangian quarks are not gauge-invariant, they do not exist. The underlying quantum field(s) cannot be excited in a way such that a single free quark is in the universe. But hadrons are gauge-invariant objects. The underlying quantum field(s) can be excited such that two or more quarks make up a composite hadron. Just like a bare quark can only exist as being part of a composite object consisting of several quarks, the respective bare quark field can only exist as being a composite of an overall gauge-invariant quantum field. Analogously, the Lagrangian electron can only exist as being part of a composite object consisting of the electron and a dressing cloud of photons. One option to try to make sense of this is by pursuing a relational approach à la \cite{Rovelli2014} as done, e.g., in \cite{Francois2023} and \cite{Gomes2024}. This could be embedded within a broader philosophical framework by engaging Rovelli's relationalism with ontic structural realism. However, all of this requires more detailed philosophical attention. Arguing that more philosophical work on this topic is to be done, is one of the main objectives of this paper.

\newpage

\section{How the mathematical structure of quantum field theory supports gauge-invariant reformulations}

It is important to note that manifestly gauge-invariant reformulations of standard particle physics are not a mere philosophical pipe dream but are physical reality. Although, admittedly, calculations become technically more challenging than in standard perturbation theory, gauge-invariant accounts can cover all parts of the Standard Model. In QED and QCD, gauge-invariant reformulations have been pioneered and specified in the works of Martin Lavelle and David McMullan (see especially \citealp{McMullan-Lavelle1997}). %As we have seen, in QED electrons are made gauge-invariant by dressing them up with photon fields and in QCD quarks are made gauge-inavariant by dressing them up with gluon fields.
%%
%\begin{quote}
%Our first motivation for dressing the Lagrangian quarks is to note that they are not gauge invariant. The gauge symmetry shows that not all fields are physically significant. To produce a gauge invariant quark field it is necessary to dress the fermions with coloured gluons. In QED electrons must also be dressed with photons, as will be discussed below, but a major difference is that photons are not electrically charged, in contrast to the colour charged gluons. (\citealp{McMullan-Lavelle1997})
%\end{quote}
%%
In the electroweak sector, Jürg Fröhlich, Giovanni Morchio, and Franco Strocchi have formulated a fully quantized gauge-invariant account of the BEH mechanism already in the early 80s (\citealp{Frohlich1980,Frohlich1981}). This approach has largely been ignored but was recently rediscovered in the work of Axel Maas (\citealp{Maas2023,Maas2020,MaasMarkl}; for an overview on gauge-invariant approaches to the BEH mechanism, see \citealp{Maas2019}). Although these three gauge-invariant mechanisms covering QED, QCD, and the electroweak sector constitute, at first sight, separate approaches, they may all be instances of a general framework called the dressing field method (see \citealp{Attard_et_al2018}; \citealp{Francois-et-al2021}; \citealp{Francois2021}; \citealp[Section 5]{Berghofer-et-al2023}). What they all have in common is that the fundamental quantities in these frameworks are all reformulated in a manifestly gauge-invariant fashion. Thereby,  they all remove gauge dependence already at the level of the field variables, as opposed to, e.g., gauge fixing as in gauge fixing the gauge field remains a fundamental quantity (see \citealp{Berghofer-et-al2023}). What they, unfortunately, also all have in common is that they are widely ignored in philosophy of physics. As stated above, it is one of the main objectives of this piece to clarify why more philosophical reflection on such approaches is in order. In this context, it should also be mentioned that one of the main implications of these gauge-invariant approaches, namely that the fundamental building blocks of reality are not point-like entities but extended objects, is expected to have empirical consequences that should be observable at some point in the not-too-distant future (\citealp{MaasJenny,Maas2023}).

Perhaps most importantly, a strong case can be made that the very mathematical structure of field theories suggests that manifestly gauge-invariant formulations are required. This is because standard approaches such as the spontaneous breaking of local symmetries, gauge fixing, and perturbation theory are all inconsistent or at least in tension with certain mathematical results. These mathematical results are Elitzur's theorem, the Gribov ambiguity, and Haag's theorem. These theorems have been known in the physics community for decades. Since gauge-invariant accounts are perfectly consistent with them, it is an interesting question of why these results have not convinced the community that standard approaches should be replaced with gauge-invariant ones. The most likely explanation is that the success of perturbation theory has convinced most physicists that a transition to practically less suitable formulations is not required. In the philosophy literature, Elitzur's theorem and the Gribov ambiguity are not often discussed or even mentioned.\footnote{Notable exceptions are, e.g., \cite{Friederich2013,Friederich2014}, \cite{Healey2007}, \cite{Rickles2008}, and \cite{Smeenk2006}.} By contrast, several works have been explicitly devoted to Haag's theorem and it has been rightly noted that it remains an open problem to explain why perturbation theory can be so successful (\citealp{EarmanFraser}). Importantly, recently it has been argued that such an explanation can be offered from the perspective of a manifestly gauge-invariant account  (\citealp{Maas2023}). In what follows, I will briefly discuss each theorem, starting with Elitzur's.

Regarding standard explanations of the BEH mechanism that invoke the spontaneous breaking of gauge symmetries, several prominent voices in philosophy of physics have rightly pointed out that it remains mysterious how the breaking of an unphysical gauge symmetry could have a physical impact on our world (see, e.g.,
\citealp{Earman2004}; \citealp{Friederich2013}; \citealp{Lyre2008}; \citealp{Smeenk2006}). However, it has been less often pointed out that this criticism is backed up by a mathematical proof. Elitzur’s theorem states that local symmetries cannot be spontaneously broken (\citealp{Elitzur1975}). This is proven for lattice QFT and expected to be true also in the continuum (see \citealp{Maas2019} for an overview).\footnote{Englert himself emphasized in his Nobel lecture that “strictly speaking there is no spontaneous symmetry breaking of a local symmetry” (\citealp[205]{Englert2014}). Interestingly, the earliest gauge-invariant reformulation of BEH goes back to Higgs himself (\citealp{Higgs1966}). See also (\citealp{Kibble1967}).} As mentioned above, the Fröhlich-Morchio-Strocchi mechanism (FMS mechanism) constitutes a gauge-invariant alternative that does not depend on dubious concepts such as gauge SSB. In other words, a gauge-invariant approach to the BEH effect is readily available that is not in tension with Elitzur's theorem.\footnote{Already in 1966 Higgs himself insisted that ``it must be possible to rewrite the theory in a form in which only gauge-invariant variables appear” (\citealp[1162]{Higgs1966}). In general, this quotation captures adequately what I mean by gauge-invariant approaches/reformulations.} There is no tension with Elitzur's theorem for the simple reason that there is no gauge SSB in FMS.

We now turn to the Gribov ambiguity. In order to quantize gauge theories, it is necessary to remove the gauge freedom. One way to do so is by virtue of manifestly gauge-invariant formulations. However, the standard and more convenient way is gauge fixing. In gauge fixing, a so-called gauge-fixing condition is specified such that precisely one element of each gauge orbit is selected (see \citealp{Boehm2001} and \citealp[58]{Rickles2008}). In other words, the space of representatives constitutes a submanifold, namely the \textit{gauge slice}, such that the slice intersects each gauge orbit exactly once. Similar to gauge SSB, gauge fixing also faces an interrelated twofold problem. On a conceptual level, gauge fixings still operate with the gauge fields as their central objects. These are gauge-variant quantities that depend on the respective gauge-fixing condition. By contrast, as mentioned above, gauge-invariant formulations remove gauge dependence already at the level of the field variables. The Gribov ambiguity thus reminds us that fixing a gauge is analogous to coordinate choices in General Relativity. There is some consensus in the foundations and philosophy of GR that the physics should not depend on our coordinate choices and the spirit of this paper is to apply the same standards to particle physics.\footnote{This received view that the physical content of GR is best represented by coordinate-free formulations, as opposed to coordinate-based formulations, has been questioned by (Wallace 2019). By contrast, the present paper suggests to take the orthodox position also seriously in particle physics, with the prospect of differential geometry serving as a unifying framework for all gauge theories.} On a technical level, it has been recognized that finding a proper gauge fixing condition is very involved. This is referred to as the \textit{Gribov ambiguity}. This denotes the fact that, especially in non-abelian gauge theories, it is often impossible to avoid the situation that the gauge slice intersects some gauge orbit either not at all or more than once (\citealp{Singer1978}). In particular, no globally well-defined gauge-fixing condition is available in QCD. Now, one could simply say that this reinforces that gauge fixing may be an effective calculational tool but that it does not successfully extract a theory's physical content\footnote{The opposite has recently been suggested by (\cite{Wallace2024}), for a response see (\cite{BerghoferFrancois}).}, and that this speaks in favor of pursuing manifestly gauge-invariant approaches instead, such as the dressing field method of which the dressing of the elecron in QED is a concrete example. It is to be noted, however, that the relationship between gauge fixings and dressings is intimate and subtle (see \cite{McMullan-Lavelle1997}, \cite{Gomes2024}, \cite{BerghoferFrancois}). In fact, the lack of a global gauge fixing condition in QCD implies that quarks cannot be globally dressed (\cite{McMullan-Lavelle1997}). Here is how Lavelle and McMullan assess this situation:

\begin{quote}
[W]e have seen that a nonperturbative
dressing would amount to the existence of a globally well defined gauge fixing
condition, which is not possible due to the Gribov ambiguity. We have sketched a proof of
this deep result — one of the very few rigorously known facts about the non-perturbative
structure of QCD. Given that physical quarks would need to be dressed, this shows that it
is not possible to construct a non-perturbative asymptotic quark field. This, we propose,
is a direct proof of quark confinement. \cite[30]{McMullan-Lavelle1997}
\end{quote}

In other words, color confinement is a consequence of the requirement that physical quantities must be gauge-invariant, and the implications of the Gribov ambiguity can only be understood in light of gauge-invariant approaches.

Now we turn in more detail to Haag’s theorem. Haag's theorem has received considerable attention in physics and philosophy. It seems to have profound conceptual implications as it ``is generally taken to show that there are severe to insurmountable mathematical difficulties modeling interactions in quantum field theory" (\citealp[1996]{Lupher2005}). What Haag's theorem shows is that the Hilbert spaces of a free field theory and an interacting one cannot be unitarily equivalent. This has been captured by the slogan: ``Haag's theorem is very inconvenient; it means that the interaction picture exists only if there is no interaction" (\citealp[166]{StreaterWightman1964}). Importantly, Haag's theorem seems to rule out perturbation theory as a mathematically consistent procedure. This is because perturbation theory requires the interaction picture, but the interaction picture seems to be ruled out by Haag's theorem. More precisely, perturbation theory requires the interaction picture in the sense that it requires that ``there is a smooth transition from an interacting (non-Abelian) gauge theory to a noninteracting (nongauge or at most Abelian) theory asymptotically" (\citealp[3, fn. 1]{Berghofer-et-al2023}; see also \citealp[165f.]{Maas2019}), which is precisely what seems to be ruled out by Haag's theorem. I say that the interaction picture and a fortiori perturbation theory ``seem" to be ruled out by Haag's theorem as there is no consensus on how to interpret it.\footnote{Recently, several works aimed at showing how the problems it seems to raise for interacting QFT may be avoided (\citealp[Section 10.5]{Duncan2012}; \citealp{EarmanFraser}; \citealp{Fraser2020}; \citealp{Koberinski2023}). One interesting claim in this context is that Haag's theorem is an ``artifact of an infinite volume idealization for relativistic systems" that ``can be avoided by modelling systems as having only finite extent" (\citealp{Koberinski2023}). This raises a number of fundamental questions regarding the role of idealizations in physics, the relationship between mathematics and reality, etc., that go beyond the scope of the present paper.} 

Be that as it may, the point is that Haag's theorem puts considerable pressure on the mathematical consistency of perturbation theory.\footnote{It has been argued that perturbation theory cannot even allow for the hydrogen atom to be a stable asymptotic state since this atom, consisting of a proton and an electron, ``does not belong to the perturbative asymptotic state space" and that Haag's theorem provides the deeper explanation for these problems (\citealp[125]{Maas2019b}; see also \citealp[165]{Maas2019}).} This, of course, raises the following crucial question: ``But why a method based on demonstrably false premises yields such `wonderfully accurate predictions' is a mystery that cries out for a resolution" (\citealp[308]{EarmanFraser}). In other words: ``There is, however, unfinished business in explaining why perturbation theory works as well as it does" (\citealp[307]{EarmanFraser}).\footnote{The approach to Haag’s theorem championed by Earman and Fraser is to abandon the interaction picture and adopt Haag-Ruelle scattering theory as a non-perturbative mathematically consistent alternative to the canonical approach. However, there are well-known problems surrounding Haag-Ruelle scattering theory. In particular, “[o]ne major drawback here is that Haag-Ruelle scattering theory is rather complicated, and still only deals with asymptotic states” (\citealp[8]{Koberinski2023}). Furthermore, it remains unclear how exactly this move to Haag-Ruelle scattering theory can explain the crucial question raised by Earman and Fraser, namely how the success of perturbation theory can be explained in light of Haag’s theorem (see \cite{Miller2018}).}

Now, what is particularly relevant in the present context, and what to my knowledge has neven been discussed in the philosophy literature, is that gauge-invariant approaches (i) avoid the problems associated with Haag's theorem and (ii) may even be in a position to explain the success of perturbation theory. We begin with the former point. Why do gauge-invariant approaches avoid the problems associated with Haag's theorem? This is because in such approaches, as exemplified by the FMS mechanism, ``fully and manifestly gauge-invariant operators are needed to construct asymptotic states" and ``this also elegantly satisfies Haag’s theorem, as the asymptotic states are no longer necessarily non-interacting elementary particles" (\citealp[180]{Maas2023}). This means we can approach Haag's theorem as follows: Haag's theorem shows that a free theory and an interacting theory cannot be unitarily equivalent. Typically, this is approached as follows: Here is a free theory and any other unitarily equivalent theory must also be a free theory, hence, there cannot be interacting theories. By contrast, we approach this in the opposite way: There is no truly free theory and physical asymptotic states are not required to be non-interacting Lagrangian particles. Accordingly, we say: Here is our manifestly gauge-invariant interacting theory and any other unitarily equivalent theory will also be an interacting one. The interaction picture and perturbation theory are ruled out insofar as they require an interacting theory to be unitarily equivalent to a free one. 

Now we turn to the second crucial virtue of pursuing gauge-invariant approaches, namely that this not only allows us to \textit{avoid} Haag's theorem, but that it also allows to \textit{explain} why perturbative methods work so well although they are mathematically inconsistent. This is because the FMS mechanism, a manifestly gauge-invariant approach to the BEH effect, predicts that due to certain properties of the Standard Model only slight deviations between a perturbative and a gauge-invariant treatment can be expected (\citealp{Maas2023}; see also \citealp[138-142]{Maas2019b}). Thus, from the gauge-invariant perspective, the success of perturbation theory is not a surprise but a \textit{prediction}.

In sum, by pursuing gauge-invariant approaches, we not only avoid the embarrassment of treating non-gauge-invariant quantities as physically fundamental objects but we thereby also make our physical mechanisms mathematically consistent (precisely by avoiding tensions with mathematical theorems such as Elitzur's and Haag's).\footnote{Of course, all this requires further analysis. In particular, the question of how gauge-invariant approaches relate to Haag's theorem deserves a paper on its own.}

\section{The explanatory power of gauge-invariant approaches}

Finally, I would like to emphasize that gauge-invariant approaches have further explanatory power that we have not yet addressed. In particular, rewriting the physical electron in a gauge-invariant fashion, as we did in Section 2, allows us to explain the Aharonov-Bohm effect (AB effect). This is the topic of this section. 

In the literature, we find three main approaches to the AB effect (see \citealp{Belot1998}; \citealp{Healey2007}; \citealp{Rickles2008}; \citealp{Berghofer-et-al2023}). Option 1 is to stick to the idea that in classical electromagnetism only the electric field $\vec{E}$ and the magnetic field $\vec{B}$ are real \textit{and} that the AB effect is explained by the magnetic field acting non-locally on the electron. By ``acting non-locally" I mean that  $\vec{B}$ has an effect on the electron even in a region where  $\vec{B} = 0$. This amounts to ``spooky" action at a distance. The virtue of this approach is that ontological commitments remain restricted to gauge-invariant quantities. In this way, determinism and observability are conserved. (In the sense that ontological commitment is restricted to quantities which are experimentally distinguishable and whose evolution is deterministic.) The shortcoming of this approach is that action at a distance is allowed.

Option 2 is to draw the conclusion that non-gauge-invariant quantities are physically real at least in the sense that they can have a physical effect. Since the vector potential $\vec{A}$ is non-zero around the electron, taking the vector potential to be a physical quantity allows explaining the AB effect without admitting action at a distance. In fact, this was the approach opted for by \cite{AharonovBohm1959} and prominently championed by Feynman in his famous \textit{Lectures on Phyiscs} (1963). However, by taking a gauge-invariant quantity to be physically real, this means that neither in theory nor via experiment we can determine which of the infinitely many gauge-related vector potentials is physically realized. This shortcoming has led many researchers to favor the third option.\footnote{It should be mentioned that a further severe shortcoming of this option is that it remains unclear whether it can really avoid non-locality even in the form of action at a distance (see \citealp[30]{Berghofer-et-al2023} and the references therein).}

Option 3 is to reconceptualize classical electrodynamics such that a new gauge-invariant but non-local object is considered fundamental. The perfect candidate for such an object seem to be the holonomies of vector potentials. This approach was pioneered by (\citealp{WuYang}) and has been promoted in philosophy most prominently and thoroughly by (\citealp{Healey2007}).
\begin{quote}
We conclude: (a) The field strength $f_{\mu \nu}$ underdescribes electromagnetism, i.e., different physical situations in a region may have the same $f_{\mu \nu}$. (b) The phase (1) [referring to $\frac{e}{\hbar c} \oint A_\mu dx^\mu$] overdescribes electromagnetism, i.e., different phases in a region may describe the same physical situation. What provides a complete description that is neither too much nor too little is the phase factor (2) [referring to exp$\big ( \frac{ie}{\hbar c} \oint A_\mu dx^\mu \big)$]. (\citealp[12]{WuYang})
\end{quote}

This phase factor is typically referred to as a holonomy. Importantly, the non-locality implied by this approach does not amount to action at a distance. As Rickles puts it:
\begin{quote}
However,
the non-locality is of a rather curious kind, for it isn’t of the ‘action-at-adistance’
variety that plagued the traditional account. The non-locality concerns
the ‘spread outness’ of the variables: they are not-localized at points. In the case of
holonomies and Wilson loops it is better to understand them as living in \textit{loop space}. (\citealp[50]{Rickles2008})
\end{quote}

 This is the main virtue of option 3: In contrast to option 2, ontological commitment is restricted to gauge-invariant quantities but in contrast to option 1, this is done in a way that avoids action at a distance. A shortcoming of option 3 seems to be that it does not really explain why this form of non-locality suddenly shows up.\footnote{For further problems of option 3, see \cite[6]{Wallace2014}.} To put it differently, option 3 feels a bit ad hoc: In classical electrodynamics, it seemed obvious that the fields  $\vec{E}$ and $\vec{B}$ are the fundamental physical quantities but then this picture is called into question by the AB effect, so we start looking for new gauge-invariant quantities. Holonomies fit well, so we say they have been the fundamental objects all along. Now, the question is whether there is an approach to the AB effect that can actually explain why this kind of non-locality appears when we move from the classical to the quantum description. More precisely, we are looking for an option such that we (i) restrict ontological commitments to gauge-invariant objects, (ii) avoid action at a distance, and (iii) are in a position to explain the AB effect in a non-ad hoc way.

Option 4: As we have seen in Section 2, the Lagrangian electron of QED is not a gauge-invariant quantity. The physical gauge-invariant electron takes the form (\ref{physicale}):
$$\psi^P (x) = D(x)\psi(x),$$
the Dirac phase factor being
$$D(x) = exp \bigg(-ie \int_x ^\infty dy_\mu A_\mu (y)\bigg).$$

Here the line integral expresses a non-local one-dimensional quantity, namely the Dirac string. This means that the physical electron is not a point-like particle but an extended object. Thus, according to option 4\footnote{The approach to the AB effect offered in (\cite{Wallace2014}) could be considered a fifth option that is gauge-invariant and local. Interestingly, it has been argued that Wallace's approach can be understood as an application of the dressing field method (\cite{Francois2019}, see also \cite[59]{Berghofer-et-al2023}.}, we can approach the AB effect as follows. Instead of understanding the physical situation as a point-like electron being affected (i) non-locally by the magnetic field, (ii) by the gauge-variant vector potential, or (iii) by a gauge-invariant non-local quantity such as the holonomies, the electron itself is an extended object that is affected by the magnetic field. Such an explanation of non-local quantum effects is summarized by Maas as follows:

\begin{quote}
In
QED, it is actually possible to transform the path integral into the variables [of the physical photon (\ref{physicalp}) and the physical electron (\ref{physicale})], i. e. physical degrees of freedom. Of course, because of the Dirac string, this will
no longer lead to a local Lagrangian. On the other hand, the Dirac string explains how
a non-local effect like the Aharonov-Bohm effect can emerge, or how Bell’s inequalities
can be understood. (\citealp[143]{Maas2019b})
\end{quote}

Let me spell out how this section relates to Section 2 and why it can be argued that by pursuing gauge-invariant approaches we can \textit{explain} the AB effect. In classical electrodynamics, it is clear that the gauge-invariant $\vec{E}$ and $\vec{B}$ are the physical fields and that the gauge-variant vector potential is not physically real. In QED, the elementary fields, i.e., the electromagnetic field $A_\mu$ and the electron-positron field $\psi$, are typically treated as the fundamental objects. However, both quantities are \textit{not} gauge-invariant. By ``making" the electron field gauge-invariant, we arrive at (\ref{physicale}) as representing the physical electron. However, this describes a non-local quantity. The physical electron can thus interact with a magnetic field even if the magnetic field is zero in the region where the point-like Lagrangian electron is located. The non-local AB effect is thus what should be expected from the gauge-invariant perspective. Importantly, this is not a retrospective ad-hoc explanation of the AB effect. This is also evident from the historical course of events, as Dirac suggested capturing the physical electron in terms of (\ref{physicale}) in 1955, several years before Aharonov and Bohm published the theoretical prediction of the AB effect (1959). Of course, option 3 and option 4 are compatible. That is, nothing prevents a proponent of gauge-invariant approaches to consider holonomies as physically real and even fundamental. 

In Section 3, we saw how pursuing gauge-invariant approaches allows us to make our physical theories consistent with the mathematical framework in which they are formulated, and in this section we discussed how this helps us to explain a physical phenomenon. It is to be noted that here we focused our discussion on QED. In the context of QCD, it has been argued by \cite{McMullan-Lavelle1997} that (i) it can be proven that color charges are only well-defined on gauge-invariant states and that pursuing a gauge-invariant account (ii) removes the infra-red problems that plague QCD and (iii) expains the phenomenon of color confinement. These results as well have been largely ignored in the philosophy (and also the physics) literature.

\section*{Conclusion}

In Section 2, we addressed the noticeable and well-motivated consensus among physicists and philosophers that only gauge-invariant quantities can be physically real. We showed that in this light it is surprising and inconsistent that the ``elementary" fields of QFT (or their excitations) are typically considered the basic physical quantities of QFT. This is problematic because these ``elementary" fields are not gauge-invariant. In Section 3, we saw how pursuing gauge-invariant approaches allows us to make our physical theories consistent with the mathematical framework in which they are formulated. This is because mathematical theorems such as Elitzur's theorem, the Gribov ambiguity, and Haag's theorem strongly suggest that standard approaches such as spontaneous breaking of gauge symmetries, gauge fixing, and perturbation theory are problematic for various reasons. All these problems can be avoided by gauge-invariant approaches. In Section 4, we addressed that pursuing gauge-invariant approaches provides us with an explanation of certain physical phenomena. Here our focus was on the Aharonov-Bohm effect. In Section 2, we introduced the physical, gauge-invariant electron of QED, and in Section 4 we saw how this helps us to make sense of the AB effect.

\section*{Acknowledgements}

I would like to thank Jordan François, Axel Maas, and Henrique Gomes for many helpful discussions and comments on an earlier version of this paper. This research was funded in part by the Austrian Science Fund (FWF) [P 36542].

\clearpage

%\printendnotes

%\clearpage

%%%%%%%%%%%%%%%%%%%%%%%%%%%%%%%%%%%%%%%%%%
%%%%%%%%%%%%%%%%%%%%%%%%%%%%%%%%%%%%%%%%%%
{
%\Huge
%\huge
%\LARGE
%\Large
%\large
\normalsize %(default)
%\small
%\footnotesize
%\scriptsize
%\tiny
 \bibliography{literature}

\begin{thebibliography}{}

\bibitem [\protect \citeauthoryear {%
Aharonov%
\ \BBA {} Bohm%
}{%
Aharonov%
\ \BBA {} Bohm%
}{%
{\protect \APACyear {1959}}%
}]{%
AharonovBohm1959}
\APACinsertmetastar {%
AharonovBohm1959}%
\begin{APACrefauthors}%
Aharonov, Y.%
\BCBT {}\ \BBA {} Bohm, D.%
\end{APACrefauthors}%
\unskip\
\newblock
\APACrefYearMonthDay{1959}{}{}.
\newblock
{\BBOQ}\APACrefatitle {Significance of Electromagnetic Potentials in the Quantum Theory} {Significance of electromagnetic potentials in the quantum theory}.{\BBCQ}
\newblock
\APACjournalVolNumPages{The Physical Review}{115}{}{485--491}.
\PrintBackRefs{\CurrentBib}

\bibitem [\protect \citeauthoryear {%
Anderson%
}{%
Anderson%
}{%
{\protect \APACyear {1967}}%
}]{%
Anderson1967}
\APACinsertmetastar {%
Anderson1967}%
\begin{APACrefauthors}%
Anderson, J.%
\end{APACrefauthors}%
\unskip\
\newblock
\APACrefYear{1967}.
\newblock
\APACrefbtitle {Principles of relativity physics} {Principles of relativity physics}.
\newblock
\APACaddressPublisher{}{Academic Press}.
\PrintBackRefs{\CurrentBib}

\bibitem [\protect \citeauthoryear {%
Attard%
, Fran\c{c}ois%
, Lazzarini%
\BCBL {}\ \BBA {} Masson%
}{%
Attard%
\ \protect \BOthers {.}}{%
{\protect \APACyear {2018}}%
}]{%
Attard_et_al2018}
\APACinsertmetastar {%
Attard_et_al2018}%
\begin{APACrefauthors}%
Attard, J.%
, Fran\c{c}ois, J.%
, Lazzarini, S.%
\BCBL {}\ \BBA {} Masson, T.%
\end{APACrefauthors}%
\unskip\
\newblock
\APACrefYearMonthDay{2018}{}{}.
\newblock
{\BBOQ}\APACrefatitle {{The dressing field method of gauge symmetry reduction, a review with examples}} {{The dressing field method of gauge symmetry reduction, a review with examples}}.{\BBCQ}
\newblock
\BIn{} J.~Kouneiher\ (\BED), \APACrefbtitle {{Foundations of Mathematics and Physics one Century After Hilbert: New Perspectives}.} {{Foundations of Mathematics and Physics one Century After Hilbert: New Perspectives}.}
\newblock
\APACaddressPublisher{}{Springer}.
\PrintBackRefs{\CurrentBib}

\bibitem [\protect \citeauthoryear {%
Bagan%
, Lavelle%
\BCBL {}\ \BBA {} McMullan%
}{%
Bagan%
\ \protect \BOthers {.}}{%
{\protect \APACyear {2000}}%
{\protect \APACexlab {{\protect \BCnt {1}}}}}]{%
Lavelle-McMullan-Bagan2000a}
\APACinsertmetastar {%
Lavelle-McMullan-Bagan2000a}%
\begin{APACrefauthors}%
Bagan, E.%
, Lavelle, M.%
\BCBL {}\ \BBA {} McMullan, D.%
\end{APACrefauthors}%
\unskip\
\newblock
\APACrefYearMonthDay{2000{\protect \BCnt {1}}}{}{}.
\newblock
{\BBOQ}\APACrefatitle {Charges from Dressed Matter: Construction} {Charges from dressed matter: Construction}.{\BBCQ}
\newblock
\APACjournalVolNumPages{Annals of Physics}{282}{2}{471 - 502}.
\PrintBackRefs{\CurrentBib}

\bibitem [\protect \citeauthoryear {%
Bagan%
, Lavelle%
\BCBL {}\ \BBA {} McMullan%
}{%
Bagan%
\ \protect \BOthers {.}}{%
{\protect \APACyear {2000}}%
{\protect \APACexlab {{\protect \BCnt {2}}}}}]{%
Lavelle-McMullan-Bagan2000b}
\APACinsertmetastar {%
Lavelle-McMullan-Bagan2000b}%
\begin{APACrefauthors}%
Bagan, E.%
, Lavelle, M.%
\BCBL {}\ \BBA {} McMullan, D.%
\end{APACrefauthors}%
\unskip\
\newblock
\APACrefYearMonthDay{2000{\protect \BCnt {2}}}{}{}.
\newblock
{\BBOQ}\APACrefatitle {Charges from Dressed Matter: Physics and Renormalisation} {Charges from dressed matter: Physics and renormalisation}.{\BBCQ}
\newblock
\APACjournalVolNumPages{Annals of Physics}{282}{2}{503 - 540}.
\PrintBackRefs{\CurrentBib}

\bibitem [\protect \citeauthoryear {%
Baker%
}{%
Baker%
}{%
{\protect \APACyear {2016}}%
}]{%
Baker2016}
\APACinsertmetastar {%
Baker2016}%
\begin{APACrefauthors}%
Baker, D.%
\end{APACrefauthors}%
\unskip\
\newblock
\APACrefYear{2016}.
\newblock
\APACrefbtitle {The Philosophy of Quantum Field Theory} {The philosophy of quantum field theory}.
\newblock
\APACaddressPublisher{}{Oxford Handbooks Online}.
\newblock
\APACrefnote{10.1093/oxfordhb/9780199935314.013.33}
\PrintBackRefs{\CurrentBib}

\bibitem [\protect \citeauthoryear {%
Belot%
}{%
Belot%
}{%
{\protect \APACyear {1998}}%
}]{%
Belot1998}
\APACinsertmetastar {%
Belot1998}%
\begin{APACrefauthors}%
Belot, G.%
\end{APACrefauthors}%
\unskip\
\newblock
\APACrefYearMonthDay{1998}{}{}.
\newblock
{\BBOQ}\APACrefatitle {{Understanding electromagnetism}} {{Understanding electromagnetism}}.{\BBCQ}
\newblock
\APACjournalVolNumPages{The British Journal for the Philosophy of Science}{49}{4}{531-555}.
\PrintBackRefs{\CurrentBib}

\bibitem [\protect \citeauthoryear {%
Berghofer%
\ \BBA {} Fran\c{c}ois%
}{%
Berghofer%
\ \BBA {} Fran\c{c}ois%
}{%
{\protect \APACyear {2024}}%
}]{%
BerghoferFrancois}
\APACinsertmetastar {%
BerghoferFrancois}%
\begin{APACrefauthors}%
Berghofer, P.%
\BCBT {}\ \BBA {} Fran\c{c}ois, J.%
\end{APACrefauthors}%
\unskip\
\newblock
\APACrefYearMonthDay{2024}{}{}.
\newblock
\APACrefbtitle {Dressing vs. Fixing: On How to Extract and Interpret Gauge-Invariant Content.} {Dressing vs. fixing: On how to extract and interpret gauge-invariant content.}
\newblock
\APACrefnote{arXiv:2404.18582}
\PrintBackRefs{\CurrentBib}

\bibitem [\protect \citeauthoryear {%
Berghofer%
\ \protect \BOthers {.}}{%
Berghofer%
\ \protect \BOthers {.}}{%
{\protect \APACyear {2023}}%
}]{%
Berghofer-et-al2023}
\APACinsertmetastar {%
Berghofer-et-al2023}%
\begin{APACrefauthors}%
Berghofer, P.%
, Fran{\c c}ois, J.%
, Friederich, S.%
, Gomes, H.%
, Hetzroni, G.%
, Maas, A.%
\BCBL {}\ \BBA {} Sondenheimer, R.%
\end{APACrefauthors}%
\unskip\
\newblock
\APACrefYear{2023}.
\newblock
\APACrefbtitle {Gauge Symmetries, Symmetry Breaking, and Gauge-Invariant Approaches} {Gauge symmetries, symmetry breaking, and gauge-invariant approaches}.
\newblock
\APACaddressPublisher{}{Cambridge University Press}.
\PrintBackRefs{\CurrentBib}

\bibitem [\protect \citeauthoryear {%
B\"ohm%
, Denner%
\BCBL {}\ \BBA {} Joos%
}{%
B\"ohm%
\ \protect \BOthers {.}}{%
{\protect \APACyear {2001}}%
}]{%
Boehm2001}
\APACinsertmetastar {%
Boehm2001}%
\begin{APACrefauthors}%
B\"ohm, M.%
, Denner, A.%
\BCBL {}\ \BBA {} Joos, H.%
\end{APACrefauthors}%
\unskip\
\newblock
\APACrefYear{2001}.
\newblock
\APACrefbtitle {{Gauge theories of the strong and electroweak interaction}} {{Gauge theories of the strong and electroweak interaction}}.
\newblock
\APACaddressPublisher{Stuttgart}{Teubner}.
\PrintBackRefs{\CurrentBib}

\bibitem [\protect \citeauthoryear {%
Dirac%
}{%
Dirac%
}{%
{\protect \APACyear {1955}}%
}]{%
Dirac1955}
\APACinsertmetastar {%
Dirac1955}%
\begin{APACrefauthors}%
Dirac, P\BPBI A\BPBI M.%
\end{APACrefauthors}%
\unskip\
\newblock
\APACrefYearMonthDay{1955}{}{}.
\newblock
{\BBOQ}\APACrefatitle {Gauge-invariant formulation of Quantum Electrodynamics} {Gauge-invariant formulation of quantum electrodynamics}.{\BBCQ}
\newblock
\APACjournalVolNumPages{Canadian Journal of Physics}{33}{}{650-660}.
\PrintBackRefs{\CurrentBib}

\bibitem [\protect \citeauthoryear {%
Dirac%
}{%
Dirac%
}{%
{\protect \APACyear {1958}}%
}]{%
Dirac1958}
\APACinsertmetastar {%
Dirac1958}%
\begin{APACrefauthors}%
Dirac, P\BPBI A\BPBI M.%
\end{APACrefauthors}%
\unskip\
\newblock
\APACrefYear{1958}.
\newblock
\APACrefbtitle {The principles of Quantum Mechanics} {The principles of quantum mechanics}\ (\PrintOrdinal{4th edn}\ \BEd).
\newblock
\APACaddressPublisher{}{Oxford University Press}.
\PrintBackRefs{\CurrentBib}

\bibitem [\protect \citeauthoryear {%
Duncan%
}{%
Duncan%
}{%
{\protect \APACyear {2012}}%
}]{%
Duncan2012}
\APACinsertmetastar {%
Duncan2012}%
\begin{APACrefauthors}%
Duncan, A.%
\end{APACrefauthors}%
\unskip\
\newblock
\APACrefYear{2012}.
\newblock
\APACrefbtitle {The Conceptual Framework of Quantum Field Theory} {The conceptual framework of quantum field theory}.
\newblock
\APACaddressPublisher{}{Oxford University Press}.
\PrintBackRefs{\CurrentBib}

\bibitem [\protect \citeauthoryear {%
Earman%
}{%
Earman%
}{%
{\protect \APACyear {2004}}%
}]{%
Earman2004}
\APACinsertmetastar {%
Earman2004}%
\begin{APACrefauthors}%
Earman, J.%
\end{APACrefauthors}%
\unskip\
\newblock
\APACrefYearMonthDay{2004}{}{}.
\newblock
{\BBOQ}\APACrefatitle {Laws, symmetry, and symmetry breaking: Invariance, conservation principles, and objectivity} {Laws, symmetry, and symmetry breaking: Invariance, conservation principles, and objectivity}.{\BBCQ}
\newblock
\APACjournalVolNumPages{Philosophy of Science}{71}{}{1227-1241}.
\PrintBackRefs{\CurrentBib}

\bibitem [\protect \citeauthoryear {%
Earman%
\ \BBA {} Fraser%
}{%
Earman%
\ \BBA {} Fraser%
}{%
{\protect \APACyear {2006}}%
}]{%
EarmanFraser}
\APACinsertmetastar {%
EarmanFraser}%
\begin{APACrefauthors}%
Earman, J.%
\BCBT {}\ \BBA {} Fraser, D.%
\end{APACrefauthors}%
\unskip\
\newblock
\APACrefYearMonthDay{2006}{}{}.
\newblock
{\BBOQ}\APACrefatitle {Haag's theorem and its impications for the foundations of quantum field theory} {Haag's theorem and its impications for the foundations of quantum field theory}.{\BBCQ}
\newblock
\APACjournalVolNumPages{Erkenntnis}{64}{}{305-344}.
\PrintBackRefs{\CurrentBib}

\bibitem [\protect \citeauthoryear {%
Elitzur%
}{%
Elitzur%
}{%
{\protect \APACyear {1975}}%
}]{%
Elitzur1975}
\APACinsertmetastar {%
Elitzur1975}%
\begin{APACrefauthors}%
Elitzur, S.%
\end{APACrefauthors}%
\unskip\
\newblock
\APACrefYearMonthDay{1975}{December}{}.
\newblock
{\BBOQ}\APACrefatitle {{Impossibility of spontaneously breaking local symmetries}} {{Impossibility of spontaneously breaking local symmetries}}.{\BBCQ}
\newblock
\APACjournalVolNumPages{Phys. Rev. D}{12}{12}{3978-3982}.
\PrintBackRefs{\CurrentBib}

\bibitem [\protect \citeauthoryear {%
Englert%
}{%
Englert%
}{%
{\protect \APACyear {2014}}%
}]{%
Englert2014}
\APACinsertmetastar {%
Englert2014}%
\begin{APACrefauthors}%
Englert, F.%
\end{APACrefauthors}%
\unskip\
\newblock
\APACrefYearMonthDay{2014}{}{}.
\newblock
{\BBOQ}\APACrefatitle {{The BEH mechanism and its scalar boson}} {{The BEH mechanism and its scalar boson}}.{\BBCQ}
\newblock
\APACjournalVolNumPages{Ann. Phys.}{526}{}{201-210}.
\PrintBackRefs{\CurrentBib}

\bibitem [\protect \citeauthoryear {%
Fran\c{c}ois%
}{%
Fran\c{c}ois%
}{%
{\protect \APACyear {2019}}%
}]{%
Francois2019}
\APACinsertmetastar {%
Francois2019}%
\begin{APACrefauthors}%
Fran\c{c}ois, J.%
\end{APACrefauthors}%
\unskip\
\newblock
\APACrefYearMonthDay{2019}{}{}.
\newblock
{\BBOQ}\APACrefatitle {{Artificial versus Substantial Gauge Symmetries: A Criterion and an Application to the Electroweak Model}} {{Artificial versus Substantial Gauge Symmetries: A Criterion and an Application to the Electroweak Model}}.{\BBCQ}
\newblock
\APACjournalVolNumPages{{Philosophy of Science}}{86}{3}{472-496}.
\PrintBackRefs{\CurrentBib}

\bibitem [\protect \citeauthoryear {%
Fran{\c c}ois%
}{%
Fran{\c c}ois%
}{%
{\protect \APACyear {2024}}%
}]{%
Francois2023}
\APACinsertmetastar {%
Francois2023}%
\begin{APACrefauthors}%
Fran{\c c}ois, J.%
\end{APACrefauthors}%
\unskip\
\newblock
\APACrefYearMonthDay{2024}{}{}.
\newblock
{\BBOQ}\APACrefatitle {The dressing field method for diffeomorphisms: a relational framework} {The dressing field method for diffeomorphisms: a relational framework}.{\BBCQ}
\newblock
\APACjournalVolNumPages{Journal of Physics A: Mathematical and Theoretical}{57}{}{}.
\PrintBackRefs{\CurrentBib}

\bibitem [\protect \citeauthoryear {%
François%
}{%
François%
}{%
{\protect \APACyear {2021}}%
}]{%
Francois2021}
\APACinsertmetastar {%
Francois2021}%
\begin{APACrefauthors}%
François, J.%
\end{APACrefauthors}%
\unskip\
\newblock
\APACrefYearMonthDay{2021}{}{}.
\newblock
{\BBOQ}\APACrefatitle {{Bundle geometry of the connection space, covariant Hamiltonian formalism, the problem of boundaries in gauge theories, and the dressing field method}} {{Bundle geometry of the connection space, covariant Hamiltonian formalism, the problem of boundaries in gauge theories, and the dressing field method}}.{\BBCQ}
\newblock
\APACjournalVolNumPages{J. High Energ. Phys.}{03, 225}{}{}.
\PrintBackRefs{\CurrentBib}

\bibitem [\protect \citeauthoryear {%
François%
, Parrini%
\BCBL {}\ \BBA {} Boulanger%
}{%
François%
\ \protect \BOthers {.}}{%
{\protect \APACyear {2021}}%
}]{%
Francois-et-al2021}
\APACinsertmetastar {%
Francois-et-al2021}%
\begin{APACrefauthors}%
François, J.%
, Parrini, N.%
\BCBL {}\ \BBA {} Boulanger, N.%
\end{APACrefauthors}%
\unskip\
\newblock
\APACrefYearMonthDay{2021}{}{}.
\newblock
{\BBOQ}\APACrefatitle {{Note on the bundle geometry of field space, variational connections, the dressing field method, \& presymplectic structures of gauge theories over bounded regions}} {{Note on the bundle geometry of field space, variational connections, the dressing field method, \& presymplectic structures of gauge theories over bounded regions}}.{\BBCQ}
\newblock
\APACjournalVolNumPages{J. High Energ. Phys.}{12, 186}{}{}.
\PrintBackRefs{\CurrentBib}

\bibitem [\protect \citeauthoryear {%
Fraser%
}{%
Fraser%
}{%
{\protect \APACyear {2020}}%
}]{%
Fraser2020}
\APACinsertmetastar {%
Fraser2020}%
\begin{APACrefauthors}%
Fraser, J.%
\end{APACrefauthors}%
\unskip\
\newblock
\APACrefYearMonthDay{2020}{}{}.
\newblock
{\BBOQ}\APACrefatitle {The Real Problem with Perturbative Quantum Field Theory} {The real problem with perturbative quantum field theory}.{\BBCQ}
\newblock
\APACjournalVolNumPages{British Journal for the Philosophy of Science}{71}{}{391-413}.
\PrintBackRefs{\CurrentBib}

\bibitem [\protect \citeauthoryear {%
Friederich%
}{%
Friederich%
}{%
{\protect \APACyear {2013}}%
}]{%
Friederich2013}
\APACinsertmetastar {%
Friederich2013}%
\begin{APACrefauthors}%
Friederich, S.%
\end{APACrefauthors}%
\unskip\
\newblock
\APACrefYearMonthDay{2013}{}{}.
\newblock
{\BBOQ}\APACrefatitle {Gauge symmetry breaking in gauge theories---in search of clarification} {Gauge symmetry breaking in gauge theories---in search of clarification}.{\BBCQ}
\newblock
\APACjournalVolNumPages{European Journal for Philosophy of Science}{3}{2}{157--182}.
\PrintBackRefs{\CurrentBib}

\bibitem [\protect \citeauthoryear {%
Friederich%
}{%
Friederich%
}{%
{\protect \APACyear {2014}}%
}]{%
Friederich2014}
\APACinsertmetastar {%
Friederich2014}%
\begin{APACrefauthors}%
Friederich, S.%
\end{APACrefauthors}%
\unskip\
\newblock
\APACrefYearMonthDay{2014}{}{}.
\newblock
{\BBOQ}\APACrefatitle {{A philosophical look at the Higgs mechanism}} {{A philosophical look at the Higgs mechanism}}.{\BBCQ}
\newblock
\APACjournalVolNumPages{J. Gen. Philos. Sci}{45}{}{335-350}.
\PrintBackRefs{\CurrentBib}

\bibitem [\protect \citeauthoryear {%
Fr\"ohlich%
, Morchio%
\BCBL {}\ \BBA {} Strocchi%
}{%
Fr\"ohlich%
\ \protect \BOthers {.}}{%
{\protect \APACyear {1980}}%
}]{%
Frohlich1980}
\APACinsertmetastar {%
Frohlich1980}%
\begin{APACrefauthors}%
Fr\"ohlich, J.%
, Morchio, G.%
\BCBL {}\ \BBA {} Strocchi, F.%
\end{APACrefauthors}%
\unskip\
\newblock
\APACrefYearMonthDay{1980}{}{}.
\newblock
{\BBOQ}\APACrefatitle {{Higgs phenomenon without a symmetry breaking order parameter}} {{Higgs phenomenon without a symmetry breaking order parameter}}.{\BBCQ}
\newblock
\APACjournalVolNumPages{Phys.Lett.}{B97}{}{249}.
\newblock
\begin{APACrefDOI} \doi{10.1016/0370-2693(80)90594-8} \end{APACrefDOI}
\PrintBackRefs{\CurrentBib}

\bibitem [\protect \citeauthoryear {%
Fr\"ohlich%
, Morchio%
\BCBL {}\ \BBA {} Strocchi%
}{%
Fr\"ohlich%
\ \protect \BOthers {.}}{%
{\protect \APACyear {1981}}%
}]{%
Frohlich1981}
\APACinsertmetastar {%
Frohlich1981}%
\begin{APACrefauthors}%
Fr\"ohlich, J.%
, Morchio, G.%
\BCBL {}\ \BBA {} Strocchi, F.%
\end{APACrefauthors}%
\unskip\
\newblock
\APACrefYearMonthDay{1981}{}{}.
\newblock
{\BBOQ}\APACrefatitle {{Higgs phenomenon without a symmetry breaking order parameter}} {{Higgs phenomenon without a symmetry breaking order parameter}}.{\BBCQ}
\newblock
\APACjournalVolNumPages{Nucl.Phys.}{B190}{}{553-582}.
\newblock
\begin{APACrefDOI} \doi{10.1016/0550-3213(81)90448-X} \end{APACrefDOI}
\PrintBackRefs{\CurrentBib}

\bibitem [\protect \citeauthoryear {%
Giddings%
\ \BBA {} Weinberg%
}{%
Giddings%
\ \BBA {} Weinberg%
}{%
{\protect \APACyear {2020}}%
}]{%
Giddings-Weinberg2020}
\APACinsertmetastar {%
Giddings-Weinberg2020}%
\begin{APACrefauthors}%
Giddings, S\BPBI B.%
\BCBT {}\ \BBA {} Weinberg, S.%
\end{APACrefauthors}%
\unskip\
\newblock
\APACrefYearMonthDay{2020}{Jul}{}.
\newblock
{\BBOQ}\APACrefatitle {Gauge-invariant observables in gravity and electromagnetism: Black hole backgrounds and null dressings} {Gauge-invariant observables in gravity and electromagnetism: Black hole backgrounds and null dressings}.{\BBCQ}
\newblock
\APACjournalVolNumPages{Phys. Rev. D}{102}{}{026010}.
\PrintBackRefs{\CurrentBib}

\bibitem [\protect \citeauthoryear {%
Gomes%
}{%
Gomes%
}{%
{\protect \APACyear {2024}}%
}]{%
Gomes2024}
\APACinsertmetastar {%
Gomes2024}%
\begin{APACrefauthors}%
Gomes, H.%
\end{APACrefauthors}%
\unskip\
\newblock
\APACrefYearMonthDay{2024}{}{}.
\newblock
\APACrefbtitle {Representational conventions and invariant structure.} {Representational conventions and invariant structure.}
\newblock
\APACrefnote{arXiv:2402.09198}
\PrintBackRefs{\CurrentBib}

\bibitem [\protect \citeauthoryear {%
Guillaud%
, Lazzarini%
\BCBL {}\ \BBA {} Masson%
}{%
Guillaud%
\ \protect \BOthers {.}}{%
{\protect \APACyear {2024}}%
}]{%
Guillaud}
\APACinsertmetastar {%
Guillaud}%
\begin{APACrefauthors}%
Guillaud, M.%
, Lazzarini, S.%
\BCBL {}\ \BBA {} Masson, T.%
\end{APACrefauthors}%
\unskip\
\newblock
\APACrefYearMonthDay{2024}{}{}.
\newblock
\APACrefbtitle {Gauge Fixing in QFT and the Dressing Field Method.} {Gauge fixing in qft and the dressing field method.}
\newblock
\APACrefnote{arXiv:2406.19937}
\PrintBackRefs{\CurrentBib}

\bibitem [\protect \citeauthoryear {%
Healey%
}{%
Healey%
}{%
{\protect \APACyear {2007}}%
}]{%
Healey2007}
\APACinsertmetastar {%
Healey2007}%
\begin{APACrefauthors}%
Healey, R.%
\end{APACrefauthors}%
\unskip\
\newblock
\APACrefYear{2007}.
\newblock
\APACrefbtitle {Gauging What's Real} {Gauging what's real}.
\newblock
\APACaddressPublisher{}{Oxford University Press}.
\PrintBackRefs{\CurrentBib}

\bibitem [\protect \citeauthoryear {%
Henneaux%
\ \BBA {} Teitelboim%
}{%
Henneaux%
\ \BBA {} Teitelboim%
}{%
{\protect \APACyear {1992}}%
}]{%
HT1992}
\APACinsertmetastar {%
HT1992}%
\begin{APACrefauthors}%
Henneaux, M.%
\BCBT {}\ \BBA {} Teitelboim, C.%
\end{APACrefauthors}%
\unskip\
\newblock
\APACrefYear{1992}.
\newblock
\APACrefbtitle {Quantization of gauge systems} {Quantization of gauge systems}.
\newblock
\APACaddressPublisher{}{Princeton University Press}.
\PrintBackRefs{\CurrentBib}

\bibitem [\protect \citeauthoryear {%
Higgs%
}{%
Higgs%
}{%
{\protect \APACyear {1966}}%
}]{%
Higgs1966}
\APACinsertmetastar {%
Higgs1966}%
\begin{APACrefauthors}%
Higgs, P\BPBI W.%
\end{APACrefauthors}%
\unskip\
\newblock
\APACrefYearMonthDay{1966}{}{}.
\newblock
{\BBOQ}\APACrefatitle {Spontaneous Symmetry Breakdown without Massless Bosons} {Spontaneous symmetry breakdown without massless bosons}.{\BBCQ}
\newblock
\APACjournalVolNumPages{Phys. Rev.}{145}{}{1156--1163}.
\PrintBackRefs{\CurrentBib}

\bibitem [\protect \citeauthoryear {%
Jenny%
, Maas%
\BCBL {}\ \BBA {} Riederer%
}{%
Jenny%
\ \protect \BOthers {.}}{%
{\protect \APACyear {2022}}%
}]{%
MaasJenny}
\APACinsertmetastar {%
MaasJenny}%
\begin{APACrefauthors}%
Jenny, P.%
, Maas, A.%
\BCBL {}\ \BBA {} Riederer, B.%
\end{APACrefauthors}%
\unskip\
\newblock
\APACrefYearMonthDay{2022}{}{}.
\newblock
{\BBOQ}\APACrefatitle {Vector boson scattering from the lattice} {Vector boson scattering from the lattice}.{\BBCQ}
\newblock
\APACjournalVolNumPages{Physical Review D}{105}{}{1-21}.
\PrintBackRefs{\CurrentBib}

\bibitem [\protect \citeauthoryear {%
Kibble%
}{%
Kibble%
}{%
{\protect \APACyear {1967}}%
}]{%
Kibble1967}
\APACinsertmetastar {%
Kibble1967}%
\begin{APACrefauthors}%
Kibble, T\BPBI W\BPBI B.%
\end{APACrefauthors}%
\unskip\
\newblock
\APACrefYearMonthDay{1967}{}{}.
\newblock
{\BBOQ}\APACrefatitle {{Symmetry breaking in non-abelian gauge theories}} {{Symmetry breaking in non-abelian gauge theories}}.{\BBCQ}
\newblock
\APACjournalVolNumPages{Phys. Rev.}{155}{}{1554-1561}.
\PrintBackRefs{\CurrentBib}

\bibitem [\protect \citeauthoryear {%
Koberinski%
}{%
Koberinski%
}{%
{\protect \APACyear {2023}}%
}]{%
Koberinski2023}
\APACinsertmetastar {%
Koberinski2023}%
\begin{APACrefauthors}%
Koberinski, A.%
\end{APACrefauthors}%
\unskip\
\newblock
\APACrefYearMonthDay{2023}{}{}.
\newblock
{\BBOQ}\APACrefatitle {{What good is Haag's no-go theorem? What axiomatic methods can teach us about particle physics}} {{What good is Haag's no-go theorem? What axiomatic methods can teach us about particle physics}}.{\BBCQ}
\newblock
\APACjournalVolNumPages{https://philsci-archive.pitt.edu/22807/}{}{}{}.
\PrintBackRefs{\CurrentBib}

\bibitem [\protect \citeauthoryear {%
Landau%
\ \BBA {} M.%
}{%
Landau%
\ \BBA {} M.%
}{%
{\protect \APACyear {1994}}%
}]{%
LL1994}
\APACinsertmetastar {%
LL1994}%
\begin{APACrefauthors}%
Landau, L\BPBI D.%
\BCBT {}\ \BBA {} M., L\BPBI E.%
\end{APACrefauthors}%
\unskip\
\newblock
\APACrefYear{1994}.
\newblock
\APACrefbtitle {The Classical Theory of Fields, fourth revised English edition} {The classical theory of fields, fourth revised english edition}.
\newblock
\APACaddressPublisher{}{Oxford University Press}.
\PrintBackRefs{\CurrentBib}

\bibitem [\protect \citeauthoryear {%
Lavelle%
\ \BBA {} McMullan%
}{%
Lavelle%
\ \BBA {} McMullan%
}{%
{\protect \APACyear {1995}}%
}]{%
McMullan-Lavelle1995}
\APACinsertmetastar {%
McMullan-Lavelle1995}%
\begin{APACrefauthors}%
Lavelle, M.%
\BCBT {}\ \BBA {} McMullan, D.%
\end{APACrefauthors}%
\unskip\
\newblock
\APACrefYearMonthDay{1995}{}{}.
\newblock
{\BBOQ}\APACrefatitle {Observables and gauge fixing in spontaneously broken gauge theories} {Observables and gauge fixing in spontaneously broken gauge theories}.{\BBCQ}
\newblock
\APACjournalVolNumPages{Physics Letters B}{347}{1}{89 - 94}.
\PrintBackRefs{\CurrentBib}

\bibitem [\protect \citeauthoryear {%
Lavelle%
\ \BBA {} McMullan%
}{%
Lavelle%
\ \BBA {} McMullan%
}{%
{\protect \APACyear {1997}}%
}]{%
McMullan-Lavelle1997}
\APACinsertmetastar {%
McMullan-Lavelle1997}%
\begin{APACrefauthors}%
Lavelle, M.%
\BCBT {}\ \BBA {} McMullan, D.%
\end{APACrefauthors}%
\unskip\
\newblock
\APACrefYearMonthDay{1997}{}{}.
\newblock
{\BBOQ}\APACrefatitle {Constituent quarks from {QCD}} {Constituent quarks from {QCD}}.{\BBCQ}
\newblock
\APACjournalVolNumPages{Physics Reports}{279}{}{1-65}.
\PrintBackRefs{\CurrentBib}

\bibitem [\protect \citeauthoryear {%
Lupher%
}{%
Lupher%
}{%
{\protect \APACyear {2005}}%
}]{%
Lupher2005}
\APACinsertmetastar {%
Lupher2005}%
\begin{APACrefauthors}%
Lupher, T.%
\end{APACrefauthors}%
\unskip\
\newblock
\APACrefYearMonthDay{2005}{}{}.
\newblock
{\BBOQ}\APACrefatitle {{Who proved Haag's theorem?}} {{Who proved Haag's theorem?}}{\BBCQ}
\newblock
\APACjournalVolNumPages{International Journal of Theoretical Physics}{44}{11}{226 - 236}.
\PrintBackRefs{\CurrentBib}

\bibitem [\protect \citeauthoryear {%
Lyre%
}{%
Lyre%
}{%
{\protect \APACyear {2008}}%
}]{%
Lyre2008}
\APACinsertmetastar {%
Lyre2008}%
\begin{APACrefauthors}%
Lyre, H.%
\end{APACrefauthors}%
\unskip\
\newblock
\APACrefYearMonthDay{2008}{}{}.
\newblock
{\BBOQ}\APACrefatitle {{Does the Higgs mechanism exist?}} {{Does the Higgs mechanism exist?}}{\BBCQ}
\newblock
\APACjournalVolNumPages{{International Studies in the Philosophy of Science}}{22}{2}{119-133}.
\PrintBackRefs{\CurrentBib}

\bibitem [\protect \citeauthoryear {%
Maas%
}{%
Maas%
}{%
{\protect \APACyear {2019}}%
{\protect \APACexlab {{\protect \BCnt {1}}}}}]{%
Maas2019}
\APACinsertmetastar {%
Maas2019}%
\begin{APACrefauthors}%
Maas, A.%
\end{APACrefauthors}%
\unskip\
\newblock
\APACrefYearMonthDay{2019{\protect \BCnt {1}}}{}{}.
\newblock
{\BBOQ}\APACrefatitle {{Brout-Englert-Higgs physics: From foundations to phenomenology}} {{Brout-Englert-Higgs physics: From foundations to phenomenology}}.{\BBCQ}
\newblock
\APACjournalVolNumPages{Prog. Part. Nucl. Phys.}{106}{}{132--209}.
\newblock
\begin{APACrefDOI} \doi{10.1016/j.ppnp.2019.02.003} \end{APACrefDOI}
\PrintBackRefs{\CurrentBib}

\bibitem [\protect \citeauthoryear {%
Maas%
}{%
Maas%
}{%
{\protect \APACyear {2019}}%
{\protect \APACexlab {{\protect \BCnt {2}}}}}]{%
Maas2019b}
\APACinsertmetastar {%
Maas2019b}%
\begin{APACrefauthors}%
Maas, A.%
\end{APACrefauthors}%
\unskip\
\newblock
\APACrefYear{2019{\protect \BCnt {2}}}.
\newblock
\APACrefbtitle {Quantum Field Theory II: Gauge Theories} {Quantum field theory ii: Gauge theories}.
\newblock
\APACaddressPublisher{}{Public Lecture Notes 2019/20}.
\PrintBackRefs{\CurrentBib}

\bibitem [\protect \citeauthoryear {%
Maas%
}{%
Maas%
}{%
{\protect \APACyear {2020}}%
}]{%
Maas2020}
\APACinsertmetastar {%
Maas2020}%
\begin{APACrefauthors}%
Maas, A.%
\end{APACrefauthors}%
\unskip\
\newblock
\APACrefYearMonthDay{2020}{}{}.
\newblock
{\BBOQ}\APACrefatitle {{The Fröhlich-Morchio-Strocchi mechanism and quantum gravity}} {{The Fröhlich-Morchio-Strocchi mechanism and quantum gravity}}.{\BBCQ}
\newblock
\APACjournalVolNumPages{SciPost Physics}{8}{51}{1-18}.
\PrintBackRefs{\CurrentBib}

\bibitem [\protect \citeauthoryear {%
Maas%
}{%
Maas%
}{%
{\protect \APACyear {2023}}%
}]{%
Maas2023}
\APACinsertmetastar {%
Maas2023}%
\begin{APACrefauthors}%
Maas, A.%
\end{APACrefauthors}%
\unskip\
\newblock
\APACrefYearMonthDay{2023}{}{}.
\newblock
{\BBOQ}\APACrefatitle {{The Fröhlich-Morchio-Strocchi Mechanism: An underestimated legacy}} {{The Fröhlich-Morchio-Strocchi Mechanism: An underestimated legacy}}.{\BBCQ}
\newblock
\BIn{} A.~Cintio\ \BBA {} A.~Michelangeli\ (\BEDS), \APACrefbtitle {Trails in Modern Theoretical and Mathematical Physics.} {Trails in modern theoretical and mathematical physics.}
\newblock
\APACaddressPublisher{}{Springer}.
\PrintBackRefs{\CurrentBib}

\bibitem [\protect \citeauthoryear {%
Maas%
, Markl%
\BCBL {}\ \BBA {} Müller%
}{%
Maas%
\ \protect \BOthers {.}}{%
{\protect \APACyear {2023}}%
}]{%
MaasMarkl}
\APACinsertmetastar {%
MaasMarkl}%
\begin{APACrefauthors}%
Maas, A.%
, Markl, M.%
\BCBL {}\ \BBA {} Müller, M.%
\end{APACrefauthors}%
\unskip\
\newblock
\APACrefYearMonthDay{2023}{}{}.
\newblock
{\BBOQ}\APACrefatitle {{Exploratory applications of the Fröhlich-Morchio-Strocchi mechanism in quantum gravity}} {{Exploratory applications of the Fröhlich-Morchio-Strocchi mechanism in quantum gravity}}.{\BBCQ}
\newblock
\APACjournalVolNumPages{Physical Review D}{107}{}{1-14}.
\PrintBackRefs{\CurrentBib}

\bibitem [\protect \citeauthoryear {%
Miller%
}{%
Miller%
}{%
{\protect \APACyear {2018}}%
}]{%
Miller2018}
\APACinsertmetastar {%
Miller2018}%
\begin{APACrefauthors}%
Miller, M.%
\end{APACrefauthors}%
\unskip\
\newblock
\APACrefYearMonthDay{2018}{}{}.
\newblock
{\BBOQ}\APACrefatitle {Haag’s Theorem, Apparent Inconsistency, and the Empirical Adequacy of Quantum Field Theory} {Haag’s theorem, apparent inconsistency, and the empirical adequacy of quantum field theory}.{\BBCQ}
\newblock
\APACjournalVolNumPages{The British Journal for the Philosophy of Science}{69}{}{801--820}.
\PrintBackRefs{\CurrentBib}

\bibitem [\protect \citeauthoryear {%
Rickles%
}{%
Rickles%
}{%
{\protect \APACyear {2008}}%
}]{%
Rickles2008}
\APACinsertmetastar {%
Rickles2008}%
\begin{APACrefauthors}%
Rickles, D.%
\end{APACrefauthors}%
\unskip\
\newblock
\APACrefYear{2008}.
\newblock
\APACrefbtitle {Symmetry, structure and spacetime} {Symmetry, structure and spacetime}.
\newblock
\APACaddressPublisher{}{Elsevier}.
\PrintBackRefs{\CurrentBib}

\bibitem [\protect \citeauthoryear {%
Rovelli%
}{%
Rovelli%
}{%
{\protect \APACyear {2014}}%
}]{%
Rovelli2014}
\APACinsertmetastar {%
Rovelli2014}%
\begin{APACrefauthors}%
Rovelli, C.%
\end{APACrefauthors}%
\unskip\
\newblock
\APACrefYearMonthDay{2014}{}{}.
\newblock
{\BBOQ}\APACrefatitle {{Why gauge?}} {{Why gauge?}}{\BBCQ}
\newblock
\APACjournalVolNumPages{Foundations of Physics}{44}{1}{91--104}.
\PrintBackRefs{\CurrentBib}

\bibitem [\protect \citeauthoryear {%
Schwartz%
}{%
Schwartz%
}{%
{\protect \APACyear {2014}}%
}]{%
Schwartz2014}
\APACinsertmetastar {%
Schwartz2014}%
\begin{APACrefauthors}%
Schwartz, M\BPBI D.%
\end{APACrefauthors}%
\unskip\
\newblock
\APACrefYear{2014}.
\newblock
\APACrefbtitle {Quantum Field Theory and the Standard Model} {Quantum field theory and the standard model}.
\newblock
\APACaddressPublisher{Cambridge}{Cambridge University Press}.
\PrintBackRefs{\CurrentBib}

\bibitem [\protect \citeauthoryear {%
Singer%
}{%
Singer%
}{%
{\protect \APACyear {1978}}%
}]{%
Singer1978}
\APACinsertmetastar {%
Singer1978}%
\begin{APACrefauthors}%
Singer, I\BPBI M.%
\end{APACrefauthors}%
\unskip\
\newblock
\APACrefYearMonthDay{1978}{}{}.
\newblock
{\BBOQ}\APACrefatitle {Some Remark on the Gribov Ambiguity} {Some remark on the gribov ambiguity}.{\BBCQ}
\newblock
\APACjournalVolNumPages{Commun. Math. Phys.}{60}{}{7-12}.
\PrintBackRefs{\CurrentBib}

\bibitem [\protect \citeauthoryear {%
Smeenk%
}{%
Smeenk%
}{%
{\protect \APACyear {2006}}%
}]{%
Smeenk2006}
\APACinsertmetastar {%
Smeenk2006}%
\begin{APACrefauthors}%
Smeenk, C.%
\end{APACrefauthors}%
\unskip\
\newblock
\APACrefYearMonthDay{2006}{}{}.
\newblock
{\BBOQ}\APACrefatitle {{The elusive Higgs mechanism}} {{The elusive Higgs mechanism}}.{\BBCQ}
\newblock
\APACjournalVolNumPages{{Philosophy of Science}}{73}{5}{487-499}.
\PrintBackRefs{\CurrentBib}

\bibitem [\protect \citeauthoryear {%
Streater%
\ \BBA {} Wightman%
}{%
Streater%
\ \BBA {} Wightman%
}{%
{\protect \APACyear {1964}}%
}]{%
StreaterWightman1964}
\APACinsertmetastar {%
StreaterWightman1964}%
\begin{APACrefauthors}%
Streater, R\BPBI F.%
\BCBT {}\ \BBA {} Wightman, A\BPBI S.%
\end{APACrefauthors}%
\unskip\
\newblock
\APACrefYear{1964}.
\newblock
\APACrefbtitle {PCT, Spin and Statistics, and All That} {Pct, spin and statistics, and all that}.
\newblock
\APACaddressPublisher{}{W. A. Benjamin, Inc.}
\PrintBackRefs{\CurrentBib}

\bibitem [\protect \citeauthoryear {%
Tong%
}{%
Tong%
}{%
{\protect \APACyear {2016}}%
}]{%
Tong2016}
\APACinsertmetastar {%
Tong2016}%
\begin{APACrefauthors}%
Tong, D.%
\end{APACrefauthors}%
\unskip\
\newblock
\APACrefYear{2016}.
\newblock
\APACrefbtitle {The Quantum Hall Effect} {The quantum hall effect}.
\newblock
\APACaddressPublisher{}{Public Lecture Notes}.
\newblock
\APACrefnote{arXiv:1606.06687}
\PrintBackRefs{\CurrentBib}

\bibitem [\protect \citeauthoryear {%
Wallace%
}{%
Wallace%
}{%
{\protect \APACyear {2014}}%
}]{%
Wallace2014}
\APACinsertmetastar {%
Wallace2014}%
\begin{APACrefauthors}%
Wallace, D.%
\end{APACrefauthors}%
\unskip\
\newblock
\APACrefYearMonthDay{2014}{}{}.
\newblock
{\BBOQ}\APACrefatitle {{Deflating the Aharonov-Bohm Effect}} {{Deflating the Aharonov-Bohm Effect}}.{\BBCQ}
\newblock
\APACjournalVolNumPages{https://arxiv.org/abs/1407.5073}{}{}{}.
\PrintBackRefs{\CurrentBib}

\bibitem [\protect \citeauthoryear {%
Wallace%
}{%
Wallace%
}{%
{\protect \APACyear {2024}}%
}]{%
Wallace2024}
\APACinsertmetastar {%
Wallace2024}%
\begin{APACrefauthors}%
Wallace, D.%
\end{APACrefauthors}%
\unskip\
\newblock
\APACrefYearMonthDay{2024}{}{}.
\newblock
\APACrefbtitle {Gauge invariance through gauge fixing.} {Gauge invariance through gauge fixing.}
\newblock
\APACrefnote{arXiv:2404.15456v1}
\PrintBackRefs{\CurrentBib}

\bibitem [\protect \citeauthoryear {%
Weinberg%
}{%
Weinberg%
}{%
{\protect \APACyear {2001}}%
}]{%
Weinberg2001}
\APACinsertmetastar {%
Weinberg2001}%
\begin{APACrefauthors}%
Weinberg, S.%
\end{APACrefauthors}%
\unskip\
\newblock
\APACrefYear{2001}.
\newblock
\APACrefbtitle {Facing Up} {Facing up}.
\newblock
\APACaddressPublisher{}{Harvard University Press}.
\PrintBackRefs{\CurrentBib}

\bibitem [\protect \citeauthoryear {%
Wu%
\ \BBA {} Yang%
}{%
Wu%
\ \BBA {} Yang%
}{%
{\protect \APACyear {1975}}%
}]{%
WuYang}
\APACinsertmetastar {%
WuYang}%
\begin{APACrefauthors}%
Wu, T\BPBI T.%
\BCBT {}\ \BBA {} Yang, C\BPBI N.%
\end{APACrefauthors}%
\unskip\
\newblock
\APACrefYearMonthDay{1975}{}{}.
\newblock
{\BBOQ}\APACrefatitle {Concept of nonintegrable phase factors and global formulation of gauge fields} {Concept of nonintegrable phase factors and global formulation of gauge fields}.{\BBCQ}
\newblock
\APACjournalVolNumPages{Physical Review D}{12}{12}{3845}.
\PrintBackRefs{\CurrentBib}

\bibitem [\protect \citeauthoryear {%
Zee%
}{%
Zee%
}{%
{\protect \APACyear {2010}}%
}]{%
Zee2010}
\APACinsertmetastar {%
Zee2010}%
\begin{APACrefauthors}%
Zee, A.%
\end{APACrefauthors}%
\unskip\
\newblock
\APACrefYear{2010}.
\newblock
\APACrefbtitle {Quantum field theory in a nutshell} {Quantum field theory in a nutshell}\ (\PrintOrdinal{2nd}\ \BEd).
\newblock
\APACaddressPublisher{}{Princeton University Press}.
\PrintBackRefs{\CurrentBib}

\bibitem [\protect \citeauthoryear {%
Zinn-Justin%
}{%
Zinn-Justin%
}{%
{\protect \APACyear {2002}}%
}]{%
ZinnJustin}
\APACinsertmetastar {%
ZinnJustin}%
\begin{APACrefauthors}%
Zinn-Justin, J.%
\end{APACrefauthors}%
\unskip\
\newblock
\APACrefYear{2002}.
\newblock
\APACrefbtitle {Quantum field theory and critical phenomena} {Quantum field theory and critical phenomena}\ (\PrintOrdinal{4th}\ \BEd).
\newblock
\APACaddressPublisher{}{Oxford University Press}.
\PrintBackRefs{\CurrentBib}

\end{thebibliography}
}

\end{document}